\documentclass[10pt,letterpaper,twocolumn]{article}
\usepackage[top=0.5in,left=0.5in,footskip=0.75in,marginparwidth=2in]{geometry}

\usepackage[utf8]{inputenc}

\usepackage{cite}

\usepackage{nameref,hyperref}

\newcommand{\mytitle}{Hunting in the Dark: Metrics for Early Stage Traffic Discovery}
\title{\mytitle}
\author{Gao, Max (UCSD)\\
magao@ucsd.edu 
\and
Collins, Michael (USC-ISI)\\
mcollins@isi.edu
\and
Mok, Ricky (CAIDA)\\
cskpmok@caida.org
\and
Claffy, kc (CAIDA)\\
kc@caida.org
}
\hypersetup{
colorlinks=true,
linkcolor=blue,
filecolor=blue,
urlcolor=blue,
citecolor=blue,
pdfpagemode=FullScreen,
pdftitle=\mytitle,
}

\usepackage[right]{lineno}

\usepackage{microtype}

\usepackage{changepage}

\usepackage[aboveskip=1pt,labelfont=bf,labelsep=period,singlelinecheck=off]{caption}

\makeatletter
\renewcommand{\@biblabel}[1]{\quad#1.}
\makeatother

\usepackage{color}
\usepackage{amssymb}
\usepackage[labelformat=parens,subrefformat=parens]{subcaption}
\usepackage{xspace}
\usepackage{multirow}
\usepackage[ruled,vlined]{algorithm2e}
\usepackage[normalem]{ulem}
\usepackage{booktabs}
\usepackage{lastpage,fancyhdr,graphicx}
\usepackage{epstopdf}
\fancyheadoffset[L]{2.25in}
\fancyfootoffset[L]{2.25in}

\definecolor{Gray}{gray}{.25}

\usepackage{sidecap}

\usepackage{wrapfig}
\usepackage[pscoord]{eso-pic}
\usepackage[fulladjust]{marginnote}
\reversemarginpar

\begin{document}
\newcommand{\etc}{{\em etc.}}
\newcommand{\ie}{{\em i.e.}}
\newcommand{\eg}{{\em e.g.}}
\newcommand{\etal}{{\em et al.}}

\newcommand{\pktset}{\ensuremath{\mathcal{P}}}
\newcommand{\pkt}{\ensuremath{p}}

\newcommand{\fspc}[1]{\ensuremath{\mathrm{#1}}}
\newcommand{\fsip}{\fspc{sip}}
\newcommand{\fdip}{\fspc{dip}}
\newcommand{\fsport}{\fspc{sp}}
\newcommand{\fdport}{\fspc{dp}}
\newcommand{\ftime}{\fspc{time}}
\newcommand{\fbytes}{\fspc{bytes}}

\newcommand{\addr}{\ensuremath{a}}
\newcommand{\addresscount}{\ensuremath{A}}
\newcommand{\cidrcount}{\ensuremath{A_{24}}}
\newcommand{\addrspread}{\ensuremath{D_\mathrm{src}}}
\newcommand{\netmask}{\ensuremath{M}}
\newcommand{\sizemask}{\ensuremath{L}}
\newcommand{\ent}{\ensuremath{S}}
\newcommand{\sizecount}{\ensuremath{G}}
\newcommand{\pktcount}{\ensuremath{C}}
\newcommand{\pktszd}{\ensuremath{Z}}
\newcommand{\disco}{\ensuremath{\mathcal{D}}}
\newcommand{\att}{\ensuremath{a}}
\newcommand{\metric}{\ensuremath{M}}
\newcommand{\rrank}{\ensuremath{\mathcal{R}}}
\newcommand{\auc}{\ensuremath{\mathcal{A}}}

\newcommand{\scanrate}{\ensuremath{r}}
\newcommand{\scannetsize}{\ensuremath{k}}
\newcommand{\probcoll}{\ensuremath{\mathcal{P}_c}}
\newcommand{\probobs}{\ensuremath{\mathcal{P}_o}}

\newcommand{\nmusc}{Group1\xspace}
\newcommand{\nmucsd}{Group2\xspace}
\newcommand{\dsusc}{\ensuremath{\mathsf{G1}}\xspace}
\newcommand{\dssd}{\ensuremath{\mathsf{G2}}\xspace}

\maketitle
\begin{abstract}
Threat hunting is an operational security process where an expert analyzes traffic, applying knowledge and lightweight tools on unlabeled data in order to identify and classify previously unknown phenomena.  In this paper, we examine threat hunting metrics and practice by studying the detection of Crackonosh, a cryptojacking malware package, has on various metrics for identifying its behavior.  Using a metric for {\em discoverability}, we model the ability of defenders to measure Crackonosh traffic as the malware population decreases, evaluate the strength of various detection methods, and demonstrate how different darkspace sizes affect both the ability to track the malware, but enable emergent behaviors by exploiting attacker mistakes.
\end{abstract}

\section{Introduction}
\label{s:intro}

Threat hunting is a proactive and situational security analysis process~\cite{collins18,badva24} in which analysts apply expertise and lightweight tools to discover malicious content.
In this paper, we examine tools used to hunt for the Crackonosh botnet, and evaluate them in terms of their {\em discoverability} over time.  
Threat hunting is a situational and adversarial process; over time, the target of the hunt changes behavior which, in turn, affects the usefulness of any particular tool.  For example, the population of any particular malware is the end result of a conflict between malware authors and multiple uncoordinated system defenders, causing the population to grow or shrink based on their actions.  These changes in population and behavior affect the efficacy of different hunting techniques, requiring that an effective hunter switch between different lightweight exploratory techniques such as clustering and stacking~\cite{uk19,auda23,collins18,szili19,ms22,zimm14,sans19,zahr21}. 

Discoverability is the probability that, when a threat hunter applies a particular metric to a dataset containing suspicious data, the cause of the suspicious data will be readily discernible.  
Discoverability follows from the intuition that analysts have limited time to examine {\em any} phenomenon and must choose the most pressing problems they face -- an analyst may be able to investigate five options in a shift, but not a hundred. 
In this paper, we evaluate and compare multiple metrics using a model of discoverability and further investigate how outside events impact detection.
Then, by comparing data against two darkspaces, one considerably larger than the other, we show how different data collection systems introduce secondary properties that can improve hunting.

We test our metrics using data from the Crackonosh~\cite{crack21} cryptojacking malware; Crackonosh targets gaming PCs by spreading through torrents containing pirated games.  
Crackonosh uses a distinct UDP-based communication scheme to update itself: every day at midnight, hosts pseudo-randomly generate a target port using a shared secret, then slowly (10 packets a second) scans the Internet on this {\em daily port} for other botnet members.  
Crackonosh is intentionally stealthy, in addition to disabling antivirus and other common host-based evasive techniques, its scanning is low, slow and originates from a small pool of sources.  
This low and slow traffic is (by design) lost in the noise at the point of origin, and will be too small to be of note for any individual honeypot.  However, Crackonosh is highly visible in darkspaces, which observe coordinated traffic to the daily port.  Crackonosh's distinct behavior means that by identifying the daily port, it is easy to extract a retrospective dataset and use it to model threat hunting. 

Within this framework, we examine the ability to monitor Crackonosh at different points in its lifetime using darkspaces. 
Since the original work on DDoS attacks by Moore~\etal~\cite{moore03,moore04}, darkspaces (also called network telescopes) have been used to examine various attacks.  
Crackonosh's scanning mechanism, relying as it does on a uniformly distributed scan of IPv4 space, is visible to darkspaces, and the larger the darkspace, the faster Crackonosh can be identified, which enables a slew of other detection and analysis techniques.  

By examining discoverability as a function of time and the concomitant change in Crackonosh's population which that entails, we show the strength of darkspace-based detection for this phenomenon.
We further show that a /16 darkspace is likely to see the entire Crackonosh botnet within the course of a day by generalizing from Moore's DDoS work to include for random internet scanning.

Our paper provides the following technical contributions: we develop the concept of {\em discoverability} to describe the situational suitability of a metric, we analyze situational factors affecting Crackonosh discoverability (darkspace size and population change due to remediation), and we compare Crackonosh to random scanning based on Moore's \cite{moore03} original darkspace model to estimate how much darkspace is needed to track similarly behaving malware.

We structure the rest of this paper as follows: \S\ref{s:prev} describes previous work on malware and traffic detection, in particular darkspace analysis and threat hunting. \S\ref{s:method} describes our methodology, and \S\ref{s:results} examines Crackonosh's \textit{discoverability} through various lightweight metrics. \S\ref{s:discussion} examines the limits of detection using darkspaces of different sizes.  \S\ref{s:conc} considers implications of the work, in particular how traffic measurement and analysis techniques can facilitate operational threat hunting needs.

\section{Related Work}
\label{s:prev}
Threat hunting is the process of proactively searching for threats within a network.  Collins~\cite{collins18} defines threat hunting as an iterative research process conducted by expert analysts within a constrained time frame, Zimmerman~\etal~\cite{zimm14} describe threat hunting as a process different from detection and response as it is focused on identifying new or previously undiscovered adversaries.  Details of threat hunting as an operational
practice are published by 
operators~\cite{uk19,szili19,ms22,auda23,rodr21}, notable is the SANS 2019~\cite{sans19} survey which lists  common threat hunting techniques.  Several threat-hunting papers in the academic community assess machine learning capabilties, without particular reference to operations~\cite{hemb21,gao21,moustafa23}.  Our work  focuses on the utility of 
threat hunting techniques
based on traffic measurement and analysis, addressing issues raised by 
recent surveys of 
threat hunters \cite{badva24,maxam24}. 

We examine our subject using darkspace traffic, which researchers have used to characterize a variety of Internet-wide security events~\cite{pang04,wustrow10,bailey05,soro19}.
These empirical insights have enabled researchers to derive models of specific phenomena, such as the DoS models developed by Moore~\etal~\cite{moore03,moore04} on which we base Crackonosh's basic models.

Additional work has examined metrics for identifying and characterizing darkspace traffic.  
Zseby~\etal~\cite{zseby14} examined entropy-based metrics for identifying aberrant darkspace traffic, focusing on IP addresses and ports. 
Other work~\cite{nychis08,zseby17,lakhina05} also examined entropy for anomaly detection.  
The idea of packet size comparisons (and entropy measures) is derived from work on application identification, notably by Collins~\etal~\cite{collins06} and Karagiannis~\etal~\cite{karag05}.

Of note is a collection of darkspace traffic classification approaches, such as the NICTER project, which has developed techniques~\cite{ban17} for identifying coordination among individual hosts in a botnet.  

Bots and other malware incorporate Internet-wide scanning with other propagation techniques, such as corrupted torrents or pirated files.  Examples include the UnixPIMINE, identified by Trend Micro~\cite{piminea,martin19}, as well as botnets, notably Mirai~\cite{antona17,wagener17} whose Internet-wide scanning was studied across multiple darkspaces.  Bou-Harb~\etal\cite{bouharb16} developed a darkspace traffic analysis capability for characterizing malware by scanning behaviors.  

\section{Methodology}
\label{s:method}
Recall from \S\ref{s:intro} that threat hunters need flexible tools to adapt to changing behavior.  To measure the effectiveness of a tool, we use a quality we call {\em discoverability}.  Discoverability is motivated by the need to optimize the workflow of a threat hunter examining an unknown phenomenon.  The hunter applies a {\em metric} to the data describing the phenomenon and creates a top-$n$ list, then investigates the elements of that list in order.  Analysts have limited attention: if $n$ is too high, they will not identify the phenomenon.  Using Crackonosh as a subject, we evaluate how discoverability changes over time for multiple traffic analysis metrics.

This section is structured as follows: \S\ref{ss:data} describes the data sets we use for analysis, while \S\ref{ss:behavior} discusses Crackonosh, its Internet-observable features, the strategies the authors took to hide its presence, and how those strategies failed.  In \S\ref{ss:remediate}, we discuss remediation and how it impacts the Crackonosh population over time, the risk it imposes on further discoveries, and the value of larger darkspaces in tracking its activity.  \S\ref{ss:exploit} discusses how to leverage larger darkspaces to identify and exploit consistent behavior.  \S\ref{ss:hypo} is a catalog of the metrics we assess for identifying Crackonosh, which we evaluate in \S\ref{s:results}.

\subsection{Data Inventory}
\label{ss:data}

\begin{figure}
\centering
\includegraphics[width=3.5in,height=2in]{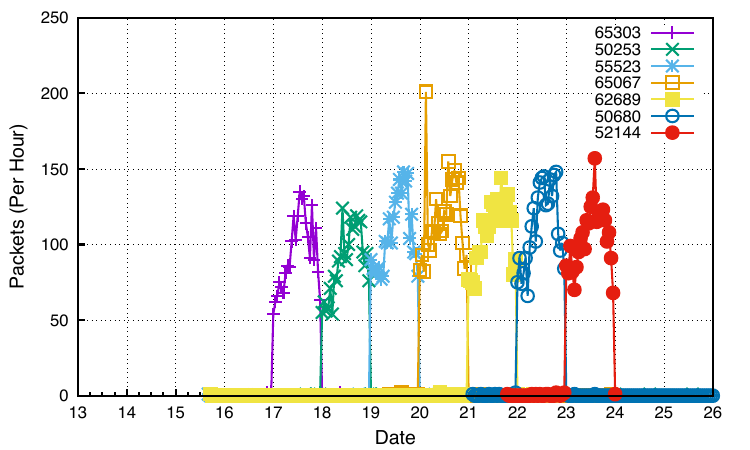}
\caption{Activity for Crackonosh ports over 2022/09/13-2022/09/26 demonstrating the coordinated rise in traffic. }
\label{f:close_ts}
\end{figure}

We used seven data sets collected from two different darkspaces, run by the two different groups contributing to this paper.  Group 1's darkspace, \dsusc, consists of a single /22. Group 2's darkspace,  \dssd, consists of 41636 /24's.  
From each darkspace, we collected data over three periods: October 13-31, 2022, January 1-15, 2024, and February 15-28, 2025.  
In addition, we used a data set from \dsusc~collected on September 13-26, 2022 for preliminary analysis (Table~\ref{t:scand} and Figure~\ref{f:close_ts}).

We labeled the data sets using a port prediction script developed by the original threat analyst who disassembled Crackonosh (see \S\ref{ss:behavior}); any UDP traffic matching the daily port is labeled as Crackonosh.  This raises a small risk of false positives where a daily port might collide with a service, however in practice this did not happen due to the high port range Crackonosh uses -- the daily port varies between 49108/UDP and 65535/UDP, while attackers focus mostly on services with much lower port numbers.

\subsection{Crackonosh and Its Observable Network Behaviors}
\label{ss:behavior}

Crackonosh is cryptojacking malware that spreads through torrents of pirated games and mines Monero using the XMRig\footnote{https://xmrig.com} cross-platform miner.  Crackonosh was initially reported in June 2021 by Avast~\cite{crack21}, a Czech antivirus developer.
Crackonosh uses multiple techniques to evade detection, including hiding control messages in encrypted DNS TXT records, disabling antivirus software, and cleaning system logs upon installation.  Crackonosh checks for updates by slowly scanning the Internet on a pseudo-randomly generated port number calculated by applying a secure hash to the date and a shared secret.

According to Avast, each infected host sends approximately 10 packets per second to random IP addresses over, while simultaneously listening on, this {\em daily port}.
A single Crackonosh host would take approximately 14 years to scan the IPv4 address space, while a network of 5,000 hosts can expect each host to contact at least one other live host per day.  In addition to the changing daily port and slow scanning, Crackonosh encrypts and pads the scan packet's payload, evading payload-based or size-based blocking.

Crackonosh is effectively a distributed daily IPv4 scan on a single port that operates stealthily enough to evade conventional scan detection -- a /22 will see any particular Crackonosh host send at most one packet a week. 
However, these same evasive behaviors distinguish Crackonosh because while most scanners focus on specific vulnerabilities, an analyst familiar with network traffic can use per-port aggregation to identify Crackonosh's unusual coordination, targets, and packet sizes.

Figure 1 shows Crackonosh's unusual coordination as observed in \dsusc~traffic from September 13 to September 26, 2022.  
Based on hourly packet counts directed to Crackonosh's daily ports, we observe two key characteristics:
1) a coordinated increase in traffic to daily ports during their active days; and 2) the relative \textit{absence} of activity on inactive days.

At 0000Z, Crackonosh will change to a new daily port, and the process repeats.

Crackonosh distinguishes itself from both opportunistic scanners and noise by its pseudo-randomly chosen daily ports which rarely intersect with ports associated with known exploits that hostile scanners commonly target.
Table I shows the top-5 busiest UDP ports by unique sender and packet counts for days between September 17 and 23, 2022.
Crackonosh's daily ports dominate the traffic, while the other ports belong to eight services with known vulnerabilities or which are used as DDoS reflectors.  We initially identified Crackonosh by noting that every day we would see a new and busy port that had no associated service.  

\begin{figure}[t]
\centering
\includegraphics[width=3.5in,height=2.5in]{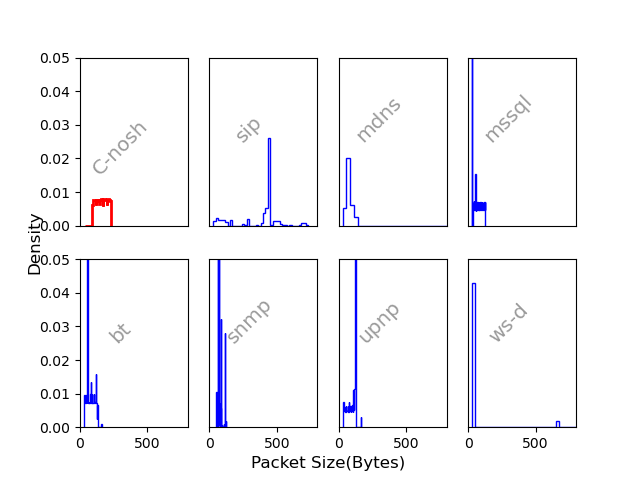}
\caption{Crackonosh's distinctive, uniform packet size distribution (red) in contrast to distributions (blue) of those targeting 8 services listed in Tab.~\ref{t:scand}.}
\label{f:bytedist}
\end{figure}

Crackonosh packets evade detection via encrypted payloads padded with a randomly determined number of bytes.  
The padding is uniformly distributed, which distinguishes it from other scan packets, which have highly modal distributions (Figure~\ref{f:bytedist}).

Group 1 and 2 initially identified Crackonosh as an oddity based on the daily port outranking other ports -- under normal circumstances, a busy UDP port has an easily searched explanation such as a vulnerability or potential as a DDoS reflector.  When ports consistently appear without associated services, this is suspicious.  
Based on these behaviors, we developed multiple metrics for discovering Crackonosh traffic, without knowing at the time what it was.
Applying these metrics to Group 2's darkspace, both teams quickly (within three hours) located daily ports and thus hosts that were potentially infected.
Infection was confirmed by a Group 2 site security team who determined individual hosts were mining Monero, which enabled both teams to definitively identify the malware and find the Avast write-up. Daniel Benes, the author of the write-up, aided us with a script to predict Crackonosh's daily port numbers. We use a variation of this script to provide ground truth ports in the G1 data set.

\begin{table*}[h]
\centering
\begin{scriptsize}
\setlength{\tabcolsep}{2.5pt}
    \begin{tabular}{c|lrr|lrr|lrr|lrr}\hline
    & \multicolumn{3}{|c|}{Sep. 17} & \multicolumn{3}{|c|}{Sep. 18} & \multicolumn{3}{|c|}{Sep. 19} & \multicolumn{3}{|c}{Sep. 20}   \\
Rank & Svc & IPs & Pkts & Svc & IPs & Pkts& Svc & IPs & Pkts& Svc & IPs & Pkts\\\hline
1 & C-nosh & 1951 & 2014  & C-nosh & 1848 & 1911   & WS-D  & 4205 & 50466 & C-Nosh & 2314& 2377\\
2 & SIP  & 913  & 33084 & WS-D & 1186 & 123058 & C-nosh & 2166 &2249 & SIP & 933 &27313\\
3 & mDNS  & 816  & 7254  & SIP & 824  & 30027  & SNMP   & 945 & 9094 & mDNS & 891 &7669\\
4 & BT  & 666  & 5503  & mDNS & 749  & 7130   & SIP  & 888 & 26925& MSSQL & 717 & 7444\\
5 & MSSQL  & 653  & 7781  & BT & 681  & 6220   & mDNS  & 782 & 7428 & BT & 708 & 5784\\\hline
& \multicolumn{3}{|c|}{Sep. 21} & \multicolumn{3}{|c|}{Sep. 22} &\multicolumn{3}{|c}{Sep. 23} & \multicolumn{3}{c}{}\\
Rank & Svc & IPs & Pkts & Svc & IPs & Pkts& Svc & IPs & \multicolumn{1}{c}{Pkts}& & & \\\cline{1-10}
1 & C-nosh & 2213 & 2279 & C-nosh & 2280  &2365 & C-nosh & 2093 &\multicolumn{1}{c}{2168} &\multicolumn{3}{c}{}\\
2 & SIP& 880& 25051 & SIP &857& 25581 & WS-D & 1378 & \multicolumn{1}{c}{45466}&\multicolumn{3}{c}{}\\
3 & mDNS & 840 & 5815 & mDNS & 840 & 8574 & SIP & 859 & \multicolumn{1}{c}{21250}& \multicolumn{3}{c}{}\\
4 & BT & 703 & 6068 & UPNP & 776 & 6634 & mDNS & 824 & \multicolumn{1}{c}{6332}&\multicolumn{3}{c}{}\\
5 & MSSQL & 652 & 7977 & BT & 701 & 5589 & BT & 700 & \multicolumn{1}{c}{6219} &\multicolumn{3}{c}{}\\\cline{1-10}

    \end{tabular}
    \caption{Busiest ports ranked by count of unique source IP addresses per day; Crackonosh's packet count is small relative to other ports, but regularly tops out the count of unique source IP's.}
    \label{t:scand}
    \end{scriptsize}
\end{table*}

\subsection{Remediation's Impact on Crackonosh's Population}
\label{ss:remediate}

\begin{figure}[h!]
	\centering
	\includegraphics[width=.45\textwidth]{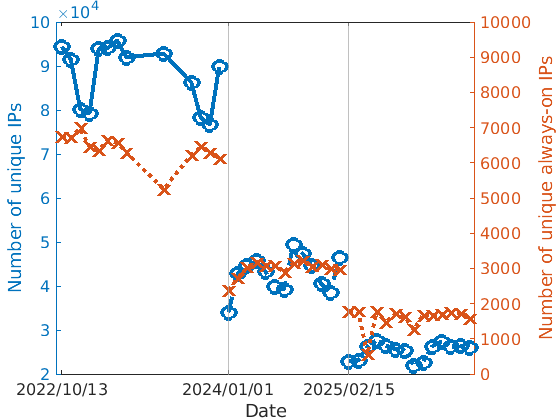}
	\caption{Number of unique IPs and always-on IPs in our data set. We removed the days that \dssd did not have complete data. (Vertical black lines represent months of missing data.) } \label{fig:datastat}
\end{figure}

Once Avast identified Crackonosh, its population steadily decreased due to remediation;  Figure~\ref{fig:datastat} shows the observed Crackonosh population in three 2-week periods across 3.5 years (October 2022, January 2024, and February 2025) captured by \dssd. This figure shows the number of addresses observed per day, which declined from $\sim$90k in 2022 to $\sim$40k in 2024, and further decreased to $\sim$26k in 2025.  Given the size of \dssd, it is reasonable to assume that these population counts represent the extant Crackonosh network. 

This population decrease makes Crackonosh progressively less discoverable using address-based metrics.  Table~\ref{t:scand} shows the busiest UDP ports, by IP address count, for September 17-23, 2022 in the \dsusc data set.  Note that in Table~\ref{t:scand}, on September 19th, 2022, Crackonosh is the {\em second} highest port by source IP address count, while {\tt ws-discovery} is the highest for that day.  The ports most commonly scanned, outside of Crackonosh, are Session Initiation Protocol (SIP, UDP/5060), Multicast DNS (mDNS, UDP/5353), BitTorrent (BT, UDP/6881), Microsoft SQL Server (MSSQL, UDP/1433), WS-Discovery (WS-D, UDP/3702), SNMP (SNMP, UDP/123), Universal Plug and Play (UPNP, UDP/1900).  All of these ports have known vulnerabilities or are used as DDoS reflectors\cite{cisa14}.  As the Crackonosh population decreases, the probability of another unrelated Internet Background Radiation (IBR) phenomenon dominating any particular metric increases.  

\subsection{Exploiting Emergent Behaviors}
\label{ss:exploit}
Using a large darkspace enables us to identify emergent phenomena, such as tracking the behavior of specific Crackonosh hosts to infer specific behaviors.  For example, we can estimate the number of packets each host sends and compare it to the results from Avast's disassembly.  To do so, we define \emph{always-on} IPs as the ones that the network telescope captured {\em at least one} probe packet from all 144 five-minute intervals within the same day. The number of \emph{always-on} IPs followed a similar declining trend: (approximately 6k in 2022, 3k in 2024, and 1.6k in 2025).
\dsusc does not observe always-on addresses, due to its smaller size. 

Using \emph{always-on} addresses, we can infer the probing rate of individual Crackonosh hosts.  Crackonosh randomly selects targets from the entire IPv4 address space, generating packets similar to backscatter resulting from randomly spoofed denial-of-service attacks (RSDoS). Therefore, we can apply the same model proposed in \cite{moore03} to estimate Crackonosh's scanning speed. Given $r$ probe packets captured by the network telescope with $\scannetsize$ IP addresses in a time interval $t$, we can estimate Crackonosh's scanning speed, $s$, with Eqn (\ref{eqn:ppstelescope}).
\begin{equation}
	s = \frac{(r / t) \times 2^{32}}{\scannetsize}. \label{eqn:ppstelescope}
\end{equation}

The main challenge in adopting this model is the availability of Crackonosh-infected hosts. Unlike RSDoS victims, which are often highly available servers, end-users may power off their machines at any time, preventing Crackonosh from sending probe packets. To obtain a more accurate estimate of $r$, we only consider the total packet counts in a day from \emph{always-on} IPs.

\begin{figure}[h!]
    \centering
    \includegraphics[width=.43\textwidth]{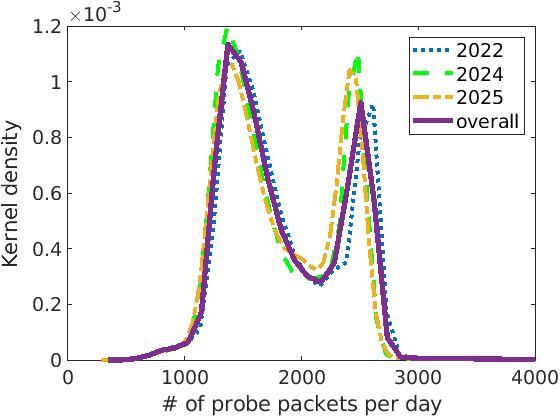}
    \caption{Kernel density function of the number of probe packets from always-on victims captured by \dssd. Bimodal distribution with peaks at 1370.31 and 2508.14 packets per day. The sending rate distribution was stable across time, and close to the expected probing rate of Crackonosh.}
    \label{fig:proberate}
\end{figure}

We employ the kernel density to infer the distribution of the total number of probe packets send by the always-on IPs in a day.
In the \dssd data sets, 65,833 unique IPs were always-on for at least one day. The kernel density of the daily probe packets captured by \dssd from these always-on hosts reveals a bimodal distribution  with similar peaks across 3.5 years (Fig. \ref{fig:proberate}). Applying Eqn (\ref{eqn:ppstelescope}), the two peaks map to 12.4  and 22.7 packets per second (pps). The lower rate aligns with the observation in \cite{crack21}, {\ie}, 10pps. The higher rate is probably due to two infected hosts behind home routers, sharing the same public IP. Furthermore, the probing rate was stable over time, showing that Cracknosh did not update this mechanism over the last few years.

\subsection{Classifying and Comparing Detection Metrics}
\label{ss:hypo}

To evaluate metrics, we estimate their \textit{discoverability} defined as $\disco_n(\pktset)$, the probability that a Crackonosh (in this case) daily port scored, under a specific metric, rank $n$ or less. 
To compute a metric's discoverability, we first partition a day's traffic by destination port number.
We then apply each of our metrics over all ports and rank the metric values, resulting in a list of (rank, port, value) tuples for each port and each day of traffic. 
From there, we compare each day's list against Crackonosh's daily port to record the daily port's corresponding rank within the list.

We limit our evaluations to the daily top-100 ports as ranked by a metric; a heuristic that assumes an operational analyst can process approximately 12 alerts per hour in 8 hours.
We average probabilities across all days of an analysis timeframe to determine a metric's aggregate discoverability.

\begin{table*}
\centering
\begin{small}
\begin{tabular}{lcp{3in}}\hline
Name & Symbol &  Description\\\hline
Source Address Count & \addresscount & Count of unique source addresses \\
Source Block Count & \cidrcount &Count of unique /24 CIDR Blocks of source IP addresses\\
Source/Dest Address Spread & \addrspread & Ratio of source and destination addresses \\
Size Entropy & \ent & Shannon entropy of individual packet sizes\\
\bottomrule
\end{tabular}
\end{small}
\caption{Summary of Metrics Used for Analysis}
\label{t:metsum}
\end{table*}

\subsubsection{Address Based Metrics: Address Count, Block Count, Spread}
\label{ss:r1}
These three metrics reflect the population of IP addresses observed per port and are most effective when 
Crackonosh's scanner population dominates interactions with a particular port.
Since non-Crackonosh scanners who scan from entire blocks of addresses (such as /24s) can confound individual source address counts, we compensate by counting source address blocks.
This proves effective as Crackonosh tends to scan from at most two addresses in a /24 network.

We calculate $\addresscount$, the address count, from a sequence of IP addresses, $\addr_0 \ldots \addr_n$, where each IP address contacts a port $p$ at least once during an observation period.
We denote the $n$-bit address count $\addresscount_{n}(\pktset, p)$ as the number of unique IP address prefixes of $n$ bits that contact port $p$. 
We refer to $\addresscount_{32}$ as the {\em address count} and $\addresscount_{24}$ as the {\em block count}.  

{\em Source address spread}, $\addrspread({\pktset,p})$, is, for a given port $p$, the ratio of source (external) addresses to destination (internal) addresses.  
The intuition behind this metric is that clients, servers, scanners and other behaviors have different and distinct ratios. For example, most scanners scan complete netblocks from a single address, resulting in a low source address spread. Crackonosh has a high source spread relative to typical scanning due to the low scan rate of individual hosts.

\subsubsection{Packet Size Metrics: Entropy}
\label{ss:r4}
The intuition behind packet size entropy as a detector is that Crackonosh's packet sizes are padded to a uniform distribution whereas the packet sizes for other probes are highly modal (Figure ~\ref{f:bytedist}). Entropy is a common anomaly detection tool ~\cite{lakhina05,nychis08,zseby14,zseby17}, although packet size itself is rarely used compared to values such as addresses.  
Crackonosh's uniformly distributed packet size results in a high entropy of between 6.8 and 7 bits.  
This entropy is considerably higher than the other protocols (Figure~\ref{f:bytedist}).  
We note that entropy exploits what we assume to be a {\em mistake} made by the Crackonosh authors; if they had not padded their payload, the resulting entropy would be smaller.

\section{Results: Comparing Discoverability Over Time}
\label{s:results}
We now compare the discoverability of Crackonosh using these metrics.  To do so, we calculate the rank and score of each metric using the \dsusc~data set across the three sample periods of October 2022, January 2024 and February 2025.  The remainder of this section is structured as follows: \S\ref{ss:resaddr} compares the three address-based metrics and \S\ref{ss:resent} examines packet size entropy.  Finally, \S\ref{ss:spped} describes the effect that different dark space sizes have on the detection time.

\subsection{Results Across Address-Bassed Metrics}
\label{ss:resaddr}
\begin{figure*}[h!]
    \begin{subfigure}[b]{\textwidth}
        \centering
        \includegraphics[width=5in]{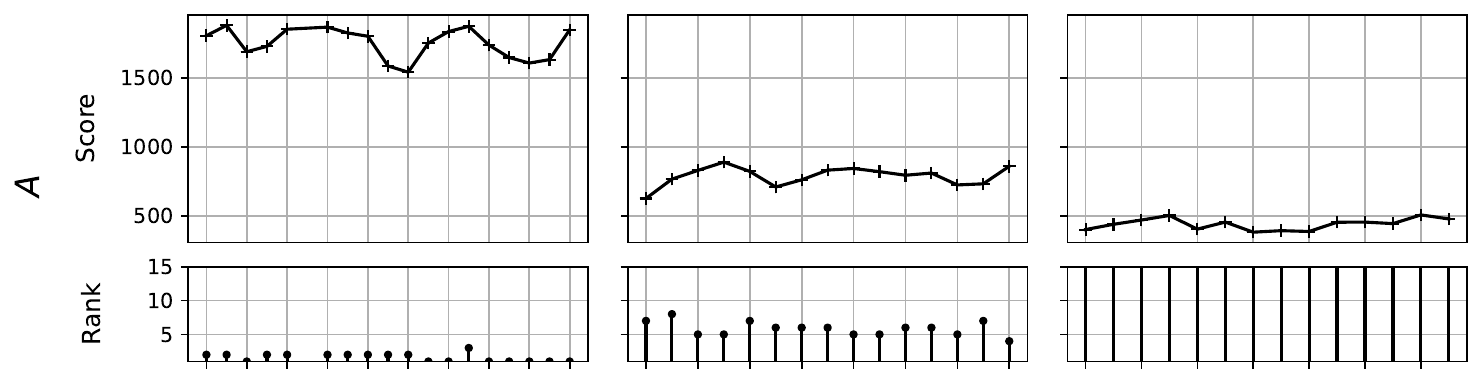}
    \end{subfigure}
    \begin{subfigure}[b]{\textwidth}
        \centering
        \includegraphics[width=5in]{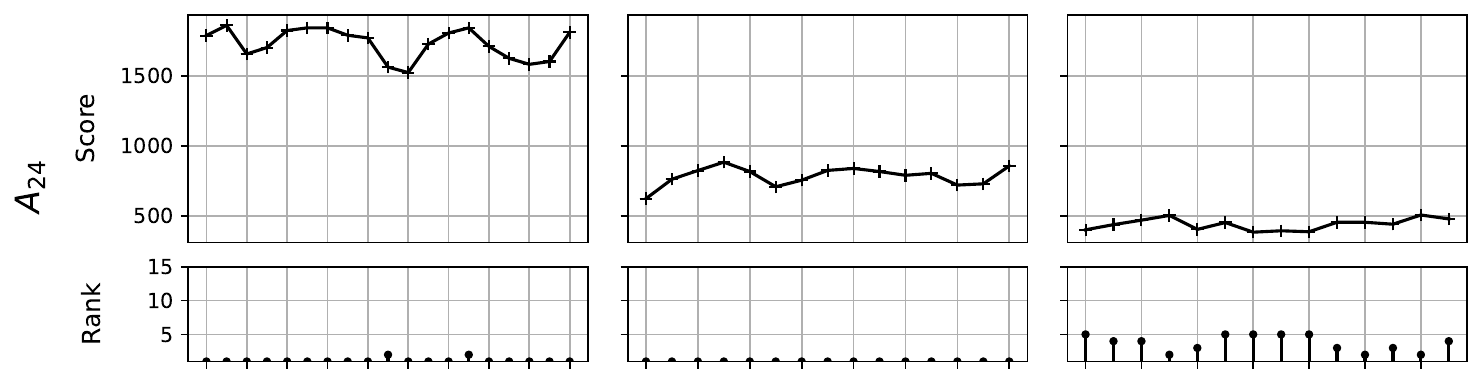}
    \end{subfigure}
    \begin{subfigure}[b]{\textwidth}
        \centering
        \includegraphics[width=5in]{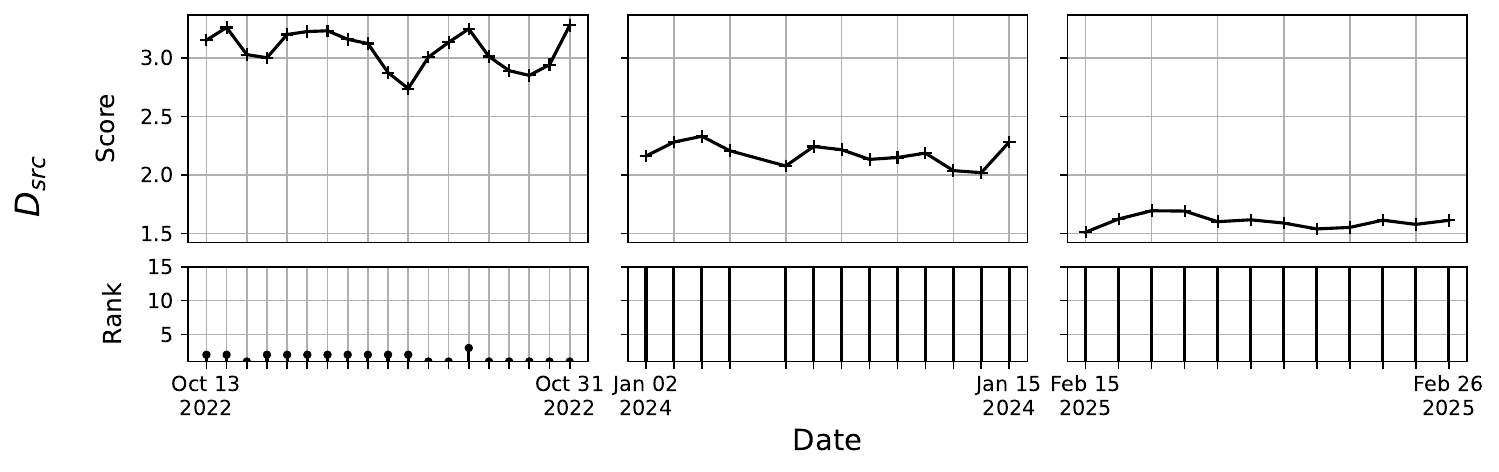}
    \end{subfigure}
    % \caption{Rank and score of Address-Based Metrics across three periods ; while the scores are correlated, the ranks differ, indicating the role of bulk scanning}
    \caption{Rank and scores of address-based metrics applied to G1's dataset across three periods. 
    While scores are correlated, the increasing ranks show the impact of other IBR obfuscating Crackonosh's behavior as the population decreases.}
    \label{f:summips}
\end{figure*}

Figure~\ref{f:summips} shows the rank and score for the address-based metrics: address count, block count and source IP spread.  Each metric is plotted across the duration of the \dsusc datasets, with lines demarcating the three periods and the corresponding date at the bottom of the plot.  Each plot in Figure~\ref{f:summips} consists of two trellised subplots -- the top is the score for each day, the bottom plot is the rank.  First, note that the three attributes are highly correlated, the Pearson Coefficient is 0.944 between address count and block count, 1 between address count and source IP spread, and 0.944 between block count and source IP address spread.  Second, despite the correlation, the {\em rank} of the daily port metric increases as the population decreases across all three metrics.  The address count and spread are invisible by 2025: at this point, the ranks regularly exceed twenty for both metrics.  In comparison,  the block count is still a viable metric, in particular because as indicated by Table~\ref{t:scand}, the ports with lower rank than Crackonosh's will be common and repeatedly seen scan targets.

The false positives in these address-based metrics are scanners, in particular scanning for UDP-based DDoS reflector ports such as SIP or mDNS (Table~\ref{t:scand}).  As Crackonosh's population dropped over the course of remediation, the probability that another scan would dominate the metric increases.  

\subsection{Results For Entropy}
\label{ss:resent}

\begin{figure*}[h]
    \centering
    \includegraphics[width=5in]{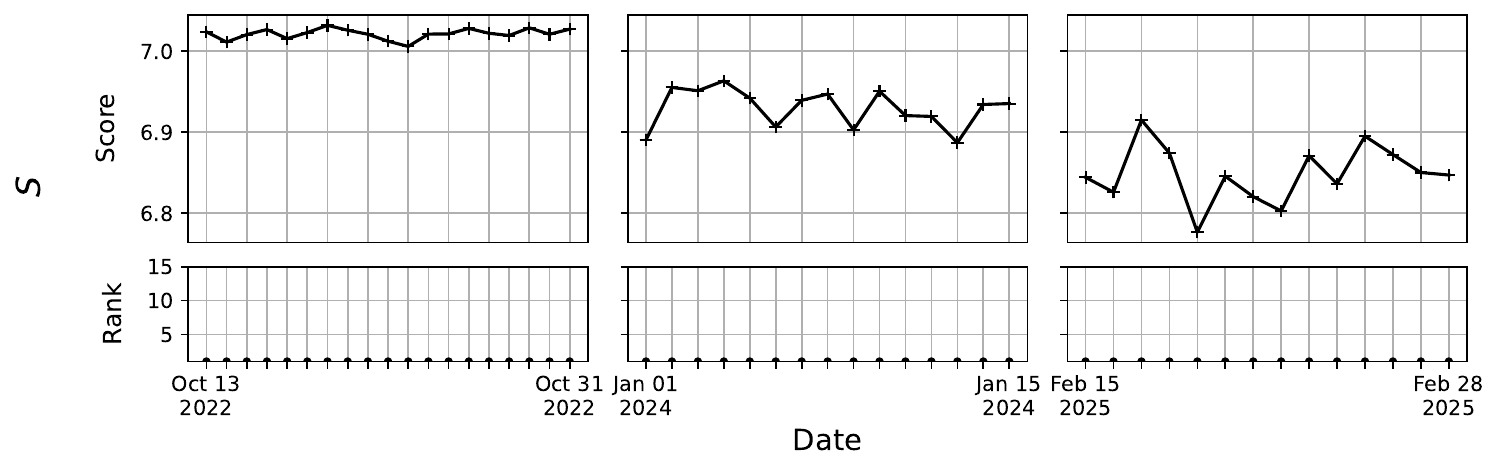}
    \caption{Compared to address-based metrics (Fig.~\ref{f:summips}), packet size entropy consistently discovers Crackonosh across each observation period.  Note that the entropy  does decrease due to less activity and fewer observed packet sizes.}
    \label{f:summpkts}
\end{figure*}

Figure~\ref{f:summpkts} summarizes the rank and score for the
entropy metric. These results are particularly notable for their consistency and high score, which we attribute to an oversight (mistake) by the attacker.
As noted in \S\ref{ss:behavior}, this uniform distribution is different from the modal packet size distributions observed for other UDP-based scanning.  
Entropy is consequently a consistently strong detector, although this metric could be easily thwarted if the attacker did not pad the packets. 
\subsection{Detection Speed}
\label{ss:spped}
Figures~\ref{f:summips} and \ref{f:summpkts} show that an analyst can reliably identify Crackonosh, even on a small darknet, within 24 hours, although by that point activity  will move onto a new daily port. The more relevant question for threat hunters is {\em how long} operators need to collect data before identifying Crackonosh or an equivalent unknown phenomenon.  

\begin{figure*}
    \centering
    \begin{subfigure}[b]{0.45\textwidth}
        \includegraphics[width=3in]{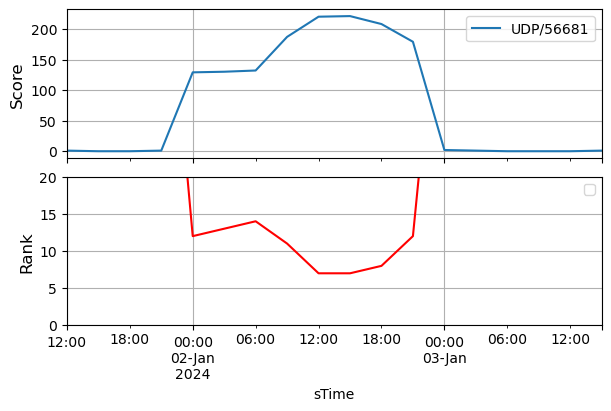}
        \caption{Group 1}
        \label{f:g1_bc}
    \end{subfigure}
    \begin{subfigure}[b]{0.45\textwidth}
        \includegraphics[width=3in]{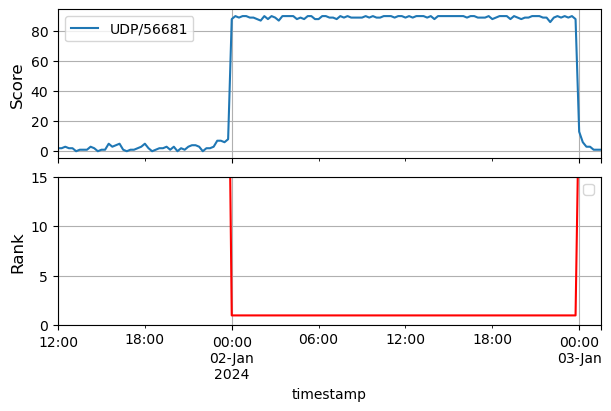}
        \caption{Group 2}
        \label{f:g2_bc}
    \end{subfigure}
    \begin{subfigure}[b]{0.45\textwidth}
        \includegraphics[width=3in]{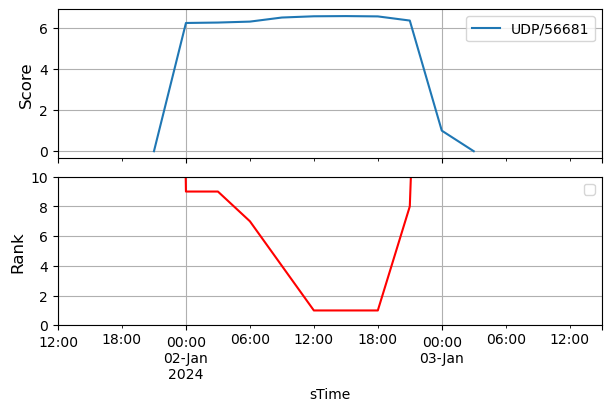}
        \caption{Group 1}
        \label{f:g1_ent}
    \end{subfigure}
    \begin{subfigure}[b]{0.45\textwidth}
        \includegraphics[width=3in]{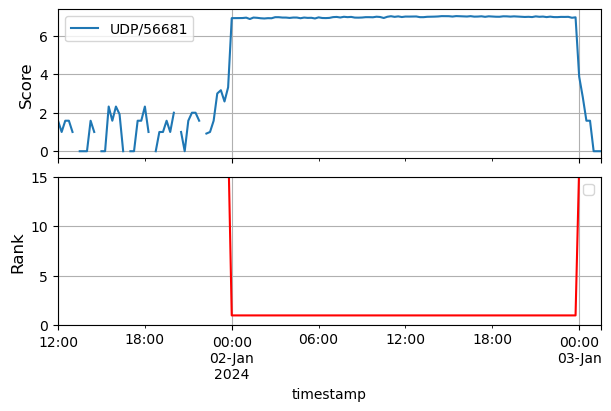}
        \caption{Group 2}
        \label{f:g2_ent}
    \end{subfigure}
    \caption{
    Comparative scores and ranks for block count (a, b) and entropy (c, d) metrics applied to 
    \nmusc and \nmucsd datasets show higher confidence and faster detection for larger darkspaces.
    }
    \label{f:bc_ent_comp}
\end{figure*}

To evaluate detection speed, we consider the impact that darkspace size has on collection and response time by applying block count (Figure~\ref{f:g1_bc},~\ref{f:g2_bc}) and entropy (Figure~\ref{f:g1_ent},~\ref{f:g2_ent}) metrics to \dsusc and \dssd darkspaces.
Figures~\ref{f:g2_bc},~\ref{f:g2_ent} plot resulting scores and ranks using 15-minute individual samples from \nmucsd~data while Figures~\ref{f:g1_bc},~\ref{f:g1_ent} use 3-hour samples from the \nmusc data.
As these figures show, the /16 used by \nmucsd~ produces top-ranked values within 15-minutes of collection, while the 3-hour sampling used by the smaller \nmusc space requires several hours to reach the same level of confidence.

\begin{figure}
    \centering
    \includegraphics[width=3in]{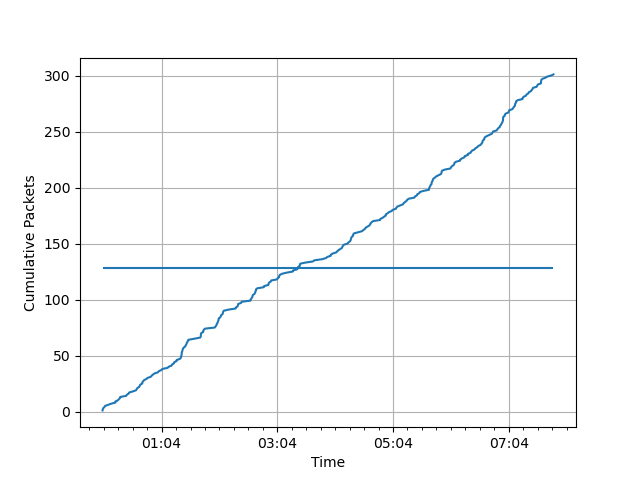}
    \caption{
    Crackonosh packet accumulation over January 2024 in the \dsusc dataset. The horizontal line indicates the minimum number of packets required to calculate entropy, equal to a time of 3 hours for a smaller-sized /22 darkspace.
    }
    \label{f:dtime}
\end{figure}

This difference raises the question of how to estimate the minimum time required to estimate the daily port.  As an initial estimate, we consider the requirements to calculate entropy. Calculating a 7-bit value for entropy of the packet size distribution requires {\em at least} 128 packets not accounting for repeated packet size.  Figure~\ref{f:dtime} plots the accumulation of Crackonosh packets from 0000Z in \dsusc-2 during January 2024.  The horizontal line in this plot indicates the 128-packet minimum needed to calculate the entropy value observed in this work.  As this figure shows, by January 2024, the time to collect the required packets is over 3 hours -- a value which is further supported by the observed score in Figure~\ref{f:g1_ent}, where a 3-hour sample still results in a score below the observed values over time.

\section{Discussion: Opportunity and Detection Limits}
\label{s:discussion}
The results from \S\ref{s:results} show how situational qualities such as remediation and darkspace size affect threat hunting.  Now, we consider the {\em limits} of detection: due to IPv4 exhaustion, large darkspaces are now rare, and we must ask at which point a darkspace becomes too small to effectively detect a phenomenon such as Crackonosh.  To do so, we will modify the original DDoS backscatter models developed by Moore~\etal~\cite{moore03,moore04} to determine when a darkspace will see too few packets to detect Crackonosh.

Moore developed models for DDoS attacks which assume that the attackers contact their target using source IP addresses uniformly spoofed across IPv4 space.  
When these packets are rejected by the target, the responses are sent to the spoofed addresses, and a darkspace has a probability---as a function of the size of the darkspace and the volume of packets sent in the attack---to receive some number of these packets.  
This type of attack results in the observing darkspace seeing a sequence of TCP packets, with a single source IP address (the target), randomly distributed IP addresses, a single source port (the targeted server), and an unknown number of destination ports (depending on whether the attacker opted to spoof their source port or not).  
In comparison to a DDoS attack, Crackonosh scanning is distributed across many sources who do not spoof their addresses, but choose a common destination port. This behavior results in the observing darkspace seeing a sequence of UDP packets, with multiple source IP addresses, randomly distributed destination IP addresses, randomly selected source ports (ephemeral UDP ports), and a single destination port.

We modify the original model by keeping in mind that Crackonosh source IP addresses are not spoofed while assuming a Crackonosh host scans IPv4 address space at random with a scanning speed $s$, targeting the  same destination port for some period $d$.  
For a single packet, the {\em probability of collision} (\probcoll) is the ratio of the collecting network size (\scannetsize) to IPv4: $\probcoll \equiv \scannetsize/2^{32}$.  
We can then define the probability of observation, $\probobs$, as the probability that {\em at least one packet}, out of a set of $sd$ packets, collides with the observed network, as follows:
\begin{equation}
\probobs \equiv 1 - (1-\probcoll)^{sd}
\end{equation}

For a single Crackonosh host, the expected number of packets sent to a single darkspace in a 24 hour period (and therefore with the same port number) is: 
\begin{equation}
    E(P) = \probcoll \cdot sd 
\end{equation}

Given a Crackonosh network of size $k$, the expected number of hosts observed by a darkspace is $k\probobs$, and the expected number of packets sent is $kE(P)$. Table~\ref{t:modresults} shows the estimated values for a /32, a /24, a /22 (\nmusc's network space), and a /16, these values are calculated with $k=1$, $s=10$ and $d=86400$.
\begin{table}
\centering
\begin{tabular}{c|ccc}
\hline
Size & \probcoll & \probobs & $E(P)$\\
\hline
/32 & 2.33E-10 & 2.01E-04 & 2.01E-04 \\
/24 & 5.96E-08 & 5.02E-02 & 5.15E-02 \\
/22 & 2.38E-07 & 0.19  & 0.21 \\
/16 & 1.53e-05 & 1.00  & 13.2 \\
\hline
\end{tabular}
\caption{Expected Crackonosh parameters as a function of darkspace size.}
\label{t:modresults}
\end{table}

The darkspace size, as shown by Table~\ref{t:modresults} demonstrates the strong impact that collector size has on the number of packets collected.  Note, in particular, that $\probobs$ for a /16 is 1.0 while the $\probobs$ for \dsusc's projected network is 0.19; by way of comparison $\probobs$ for a /18 is 0.96, and a /19 is 0.81.  These probabilities indicate that a /16 can expect to observe {\em at least} one packet from each Crackonosh host daily, while a /22 will require 13 days to reach 95\

The difference in darkspace size, and the consequent time to detect Crackonosh activity within the hard limit imposed by the port change, means that larger darkspaces introduce emergent effects.  
In particular, there are issues of maturation and attrition~\cite{shadish02}, the impact that the change in population has on detection.  
Crackonosh's population changes over time due to external (to the observer) remediation.  
Attempts to estimate the population over extended periods, such as capture-recapture techniques, must account for these population changes.  By the time enough samples are gathered to make an estimate on \dsusc, the population will have shrunk due to attrition, while one can sample \nmucsd~data for a much shorter time with less need to account for such changes.

\section{Conclusion}
\label{s:conc}
In this paper we summarized our investigation of a collection of lightweight traffic measurement and analysis metrics to identify traffic generated by the Crackonosh botnet.  Our motivation for doing so was to formalize how operational security personnel begin with an anomaly in traffic data and perform analysis to positively identify a threat.  We have done so by creating a new gauge, \textit{discoverability}, to evaluate how well a set of metrics facilitate discovery of malicious behavior over time.  

Network traffic measurement to support security operations
often involves multiple organizations with competing goals. In addition to the initial attacker and defender, there are other remediators, other attackers, and gray hat organizations that all simultaneously affect traffic. Identifying novel malicious behavior often requires exploiting specific situations.  We have formalized discoverability to account for these dynamics.  For example, Crackonosh hosts are thinly spread across networks while scanners are more tightly concentrated, meaning that Crackonosh is more discoverable using a block count rather than a simple address count.  Crackonosh is highly discoverable using entropy of the packet size distribution, but the attacker could have eliminated that problem by using modal padding or no padding.  This {\em opportunism} means that defenders must cultivate options, including a variety of metrics, collection systems, and datasets.

\appendix

\textbf{Ethics.}  The data used for this experiment consists of
unsolicited traffic sent to IPv4 darkspaces from across the
Internet. While this traffic was directed to darkspaces, it originated
from hosts infected by malware. To protect the privacy of these hosts,
the data set is available on request and subject an acceptable use
policy.

\textbf{Acknowledgements} This material is based on research sponsored
by the National Science Foundation (NSF) grants OAC-2319959 and
CNS-2120399, as well as the Amateur Radio Digital Communications
Foundation.  The views and conclusions contained herein are those of
the authors and should not be interpreted as necessarily representing
the official policies or endorsements, either expressed or implied, of
funding agencies.

\bibliographystyle{plain}
\bibliography{arxiv}
\end{document}